\def\gsim{\;\raisebox{-.4ex}{\rlap{$\sim$}} \raisebox{.4ex}{$>$}\;}
\def\lsim{\;\raisebox{-.4ex}{\rlap{$\sim$}} \raisebox{.4ex}{$<$}\;}

\documentclass[12pt]{article}
\begin{document}

\begin{flushright}
hep-ph/9808386\\
\end{flushright}

\begin{center}
{\Large\bf Phenomenology of Neutral $D-$meson Decays\\
              and Double-Flavor Oscillations
\footnote{Published in Eur.Phys.J. {\bf A4} (1999) 21-31}
}

\vspace{0.4cm}

{\large Ya.I.Azimov \footnote{e-mail: azimov@pa1400.spb.edu}}

\vspace{0.3cm}

{\it Petersburg Nuclear Physics Institute,\\
Gatchina, St.Petersburg, 188350, Russia}

\vspace{0.3cm}
\end{center}

\begin{abstract}
Decays of neutral $D$-mesons are considered phenomenologically
without invoking any particular models. Special attention is
given to cascade decays with intermediate neutral kaons where
coherent double-flavor oscillations (CDFO) become possible.
We show necessity and unique possibilities of experiments
on CDFO. They allow to relate with each other widths and
masses of $D$-meson eigenstates, to separate interference
effects due to $D^0$-$\overline D^0$ mixing and/or
Cabibbo-favored vs. doubly-suppressed transitions. Such
experiments provide the only known ways to unambiguous
model-independent measurements of all $CP$-violating parameters
and of Cabibbo-doubly-suppressed amplitudes, where the New
Physics may have more prominent manifestations. Similar
experiments would be useful and interesting also for charged
$D$-meson decays to neutral kaons.

\vspace{0.3cm}
\end{abstract}

\section {Introduction}

Study of coherent double-flavor oscillations (CDFO) was suggested
some years ago~\cite{az1} as a method for detailed investigations
of properties of heavy mesons. The phenomenon emerges if a secondary
neutral kaon produced in decay of a heavier neutral flavored meson
evolves so as to coherently continue the pre-decay evolution of the
initial heavy meson. It has been discussed in a number of papers
[1-7], mainly for $B$-mesons. The method suggests new tools to
measure $\Delta m$ and $\Delta\Gamma$ for $B_d$~\cite{az1,ds} and
$B_s$~\cite{ad} mesons, providing, in particular, a unique
possibility to find their signs. For experimental studies of
$CP$-violation it can present a practical way to measure
$CP$-violating parameters in neutral $B$-meson decays unambiguously
and independently of any model-based assumptions~\cite{ars,az2,bk}.
Detailed discussion of these and other aspects of CDFO in $B$-decays
may be found in the review talk~\cite{az3}. The problem of
ambiguities for parameters of $CP$-violation has recently been
discussed also in a large number of papers (see, e.g., refs.[9-14]).

The present paper concerns with special features of $D$-meson decays
which appear to be, in some sense, phenomenologically more general
than $B$-meson ones. $D$-meson physics has many interesting problems
(see, e.g., the detailed mini-review~\cite{gol}). One of essential
phenomenological differences between $B$- and $D$-physics is that
any particular decay of $B$-mesons corresponds to a single-flavor
transition, while $D$-mesons may have various flavor transitions
in the same decay mode.  As noted in refs.~\cite{ai,by} (see also
ref.~\cite{xing}), specifics of neutral kaons as decay products may
generate unusual effects even in decays of charged (and, surely,
unmixed) $D^\pm$-mesons. Their source is that in such decays the
Cabibbo-allowed and doubly-suppressed transitions, which are just
different flavor transitions, become coherent. As a result, in
particular, the sought-for $D$-meson $CP$-violation effects become
observationally mixed with the well-studied kaon ones.

In neutral $D$-decays the mixing of $D^0$ and $\overline D^0$  opens
possibility of CDFO and leads to additional non-standard effects.
So, in analogy with $B$-meson decays considered earlier~\cite{az1},
we are interested now in cascade decays of the type \begin{equation}
D^0 (\overline D^0)\rightarrow XK^0 (\overline K^0)\,, \end{equation}
with subsequent kaon decays; $X$ is a neutral system with definite
values of spin and $CP$-parity. Our aim is to study what physical
information may be extracted from double-time distributions over
primary and secondary lifetimes $t_D$ and $t_K$.

Decays (1) are mainly induced by the quark transitions
$c\to sW^+\,,\,\,\, \bar c\to\bar sW^-$,
which produce meson transitions
\begin{equation}D^0 \to\overline K^0\,,\,\,\, \overline D^0
\to K^0\,. \end{equation}
Their final strangeness is the same as in decays of $\overline
B_d,\, B_d$ or $B_s,\,\overline B_s$, studied in
papers~\cite{az1,ad} respectively. Hence, if only transitions (2)
existed we could apply ready expressions from those papers to
describe time distributions of decays (1). However, transitions
\begin{equation}D^0 \to K^0\,,\,\,\, \overline D^0
\to\overline K^0\,,
\end{equation}
with the "wrong" final strangeness, are also possible. Being
induced by the quark cascades $c\to dW^+\,,\,\,\,W^+\to u\bar s$
and charge conjugate, they are doubly Cabibbo-suppressed.
Nevertheless, when searching for very small expected effects of
$D$-meson mixing or $CP$-violation, the interference of transitions
(2) and (3) in the secondary kaon decays should be taken into
account. Moreover, the doubly-suppressed transitions are a new kind
of manifestations of electro-weak interactions, which may reveal
some New Physics; so their studies are of independent interest.
Therefore, we begin here with exact expressions and apply smallness
assumptions only later. (Note that in particular decays of
$B$-mesons the "wrong"-strangeness transitions are practically
absent being suppressed much stronger.) At first sight, possibility
of two transitions (2) and (3) might essentially complicate time
distributions, in comparison with $B$-mesons. We will see, however,
that the complications are, in essence, not very serious. Moreover,
they open new possibilities to extract physically interesting
information from experiments.

The further presentation goes as follows. In Section~2 we give
general description of cascades initiated by decays (1) with
subsequent kaon decays. Physical content of seemingly complicated
expressions is first discussed in Section~3 for the simplified case
of exact $CP$-conservation. For the realistic case of violated
$CP$-parity we explain in Section~4 that physical identification of
neutral $D$-meson eigenstates is important to prevent ambiguities
both in measuring $CP$-violation parameters and in separating
amplitudes of flavor transitions (2) and (3). For illustration we
consider two kinds of labeling the eigenstates. In Section~4 they
are marked as being approximately $CP$-even or $CP$-odd, while in
Section~5 we label them by the heavier or lighter mass. We show
that in both cases the double-time decay distributions of cascades
(1) are necessary and sufficient to relate together various
properties of eigenstates and eliminate ambiguities from
measurements of physical quantities. To conclude we summarize the
results and briefly discuss possible strategies of experiments.

\section {General formalism}

Neutral $D$-mesons produce two eigenstates which we denote by
$D_\pm$ (the meaning of such notations is discussed
below). For simplicity we assume $CPT$ (but not $CP$) invariance.
Then the eigenstates may be written as
\begin{equation} D_\pm =p_D D^0 \pm q_D \overline D^0\,;
\end{equation}
they have definite masses and widths; simple factors $$e_\pm (t) =
\exp (-im_\pm t -\Gamma_\pm t/2)$$ describe their time evolutions.
The relations (4) are considered in many papers as definitions of
the eigenstates (for kaons, in a standard way, $K_S$ is assumed to
be $K_+$, while $K_L$ is identified with $K_-$). We emphasize,
however, that these definitions are only formal and cannot be
considered as physical definitions of eigenstates. Since
coefficients $p, q$ are not assumed to be real, one may actually
redefine phases of states so, that any prescribed eigenstate would
look as $D_+$ (for kaons one may consider $K_S$ and $K_L$ as having
the form of $K_-$ and $K_+$ respectively, by changing phases of
$K^0$ and $\overline K^0$ without changing any physical quantities).
The problem of true physical definitions for the eigenstates will
be considered below.

In analogy with ref.~\cite{az1}, we start, say, with the pure
$D^0$-state.  During the time interval $t_D$ it evolves into the
state
\begin{equation}
D(t_D) =\frac{1}{2p_D} [e_+(t_D) D_+ + e_-(t_D) D_-]\,.
\end{equation}
Decay (1) at the moment $t_D$ generates the kaon state (up to
normalization)
\begin{equation}
K(t_D;0) =\frac{1}{2p_D} \{[a^{(X)}_{+S}\, e_+(t_D)  +
a^{(X)}_{-S}\, e_-(t_D) ]K_S + [a^{(X)}_{+L}\, e_+(t_D) +
a^{(X)}_{-L}\, e_-(t_D) ]K_L \}\,, \end{equation}
where $a^{(X)}_{\pm S}, a^{(X)}_{\pm L}$ are amplitudes of decays
(1) with transitions $D_\pm\to K_{S,L}$. Evolution during the time
$t_K$ transforms it into
$$K(t_D;t_K) =\frac{1}{2p_D}\{[a^{(X)}_{+S}\,e_+(t_D) +
a^{(X)}_{-S}\,e_-(t_D)] e_S(t_K) K_S\;\;\;\;\;\;\;$$
\begin{equation}
\;\;\;\;\;\;\;~~~~+ [a^{(X)}_{+L}\, e_+(t_D) + a^{(X)}_{-L}\,
e_-(t_D)] e_L(t_K) K_L\}\,, \end{equation}
with $e_{S,L}(t) = \exp (-im_{S,L} t - \Gamma_{S,L} t/2)\,.$

Let $b_{Sf}$ and $b_{Lf}$ denote amplitudes of decays
\begin{equation}
K_{S,L}\to f\,.
\end{equation}
Then the cascade, initiated by the pure $D^0$-meson and consisting
of the primary decay (1) after lifetime $t_D$ and the secondary kaon
decay (8) after lifetime $t_K$, has the probability amplitude equal
to $$A_{D\to Xf}(t_D;t_K) =\frac{1}{2p_D}\{[a^{(X)}_{+S}\,e_+(t_D) +
a^{(X)}_{-S}\,e_-(t_D)] b_{Sf}\,e_S(t_K) \;\;\;\;\;\;\;$$
\begin{equation} \;\;\;\;\;\;\;\;\;\;~~~~~~~~+ [a^{(X)}_{+L}\, e_+
(t_D) + a^{(X)}_{-L}\, e_-(t_D)] b_{Lf}\,e_L(t_K) \}\,.\end{equation}
The amplitude for the similar cascade initiated by the pure
$\overline D^0$-meson is somewhat different. It equals
$$A_{\overline D\to Xf}(t_D;t_K) =\frac{1}{2q_D}\{[a^{(X)}_{+S}\,
e_+(t_D) - a^{(X)}_{-S}\,e_-(t_D)] b_{Sf}\,e_S(t_K)\;\;\;\;\;\;\;$$
\begin{equation} \;\;\;\;\;\;\;\;\;\;~~~~~~~~+
[a^{(X)}_{+L}\, e_+(t_D) - a^{(X)}_{-L}\,e_-(t_D)]b_{Lf}\,
e_L(t_K)\}\,.\end{equation}
Double-time distributions of these cascades may be presented in
the form
\begin{equation} W^{Xf}(t_D;t_K)=|A_{D\to
Xf}(t_D;t_K)|^2\,;~~~~ \overline W^{Xf}(t_D;t_K)=|A_{\overline D\to
Xf}(t_D;t_K)|^2\,.  \end{equation}
Their structure reminds that described in ref.~\cite{az1}. The
distributions are not factorisable. As a function of one time (say,
of $t_D$) they are linear combinations of four terms $$\exp(-\Gamma_+
t_D)\,,~~~\exp[-(\Gamma_+ +\Gamma_-)t_D/2] \cos(\Delta m_D t_D)\,,$$
$$\exp(-\Gamma_- t_D)\,,~~~ \exp[-(\Gamma_+ +\Gamma_-)t_D/2]
\sin(\Delta m_D t_D)\,.$$ Coefficients depend on $t_K$ and, in their
turn, are also linear combinations of similar terms $$\exp(-\Gamma_S
t_K)\,,~~~\exp[-(\Gamma_S +\Gamma_L)t_K/2] \cos(\Delta m_K t_K)\,,$$
$$\exp(-\Gamma_L t_K)\,,~~~ \exp[-(\Gamma_S +\Gamma_L)t_K/2]
\sin(\Delta m_K t_K)\,.$$ Physically more transparent is a different
way of describing the double-time distributions (11).  They contain
several various contributions. First of all, there are
non-interfering contributions of 4 possible cascade branches
(corresponding to various combinations of subscripts in $D_\pm\to
K_{S,L}$):  $$|2p_D|^2\,W^{Xf}_{no~int}(t_D;t_K)=|2q_D|^2\,
\overline W^{Xf}_{no~int}(t_D;t_K)
~~~~~~~~~~~~~~~~~~~~~~~~~~~~~~~~~~~~~~~~~~~~~~~~~~~$$
$$~~~~~~~~=|a^{(X)}_{+S}b_{Sf}|^2 \exp(-\Gamma_+t_D-\Gamma_St_K)
+|a^{(X)}_{+L}b_{Lf}|^2 \exp(-\Gamma_+t_D-\Gamma_Lt_K)$$
\begin{equation}
~~~~~~~~~~+|a^{(X)}_{-S}b_{Sf}|^2 \exp(-\Gamma_-t_D-\Gamma_St_K)
+|a^{(X)}_{-L}b_{Lf}|^2 \exp(-\Gamma_-t_D-\Gamma_Lt_K)\, .
\end{equation}
Then, there are single-interference contributions. They are
due to $K_{S,L}$ interference without $D_\pm$ interference
or, vice versa, due to interference of $D_\pm$ without
$K_{S,L}$ interference:
$$|2p_D|^2\,W^{Xf}_{K~int}(t_D;t_K)=|2q_D|^2\,\overline
W^{Xf}_{K~int}(t_D;t_K)
~~~~~~~~~~~~~~~~~~~~~~~~~~~~~~~~~~~~~~~~~~~~~~~~~~~$$
$$=2{\rm Re}[a^{(X)\ast}_{+L} a^{(X)}_{+S} b_{Lf}^\ast
b_{Sf}\exp(i\Delta m_K t_K)]\exp[-\Gamma_+ t_D -(\Gamma_S
+\Gamma_L)t_K/2]$$ \begin{equation} ~~+2{\rm Re}[a^{(X)\ast}_{-L}
a^{(X)}_{-S} b_{Lf}^\ast b_{Sf}\exp(i\Delta m_K t_K)]\exp[-\Gamma_
- t_D-(\Gamma_S +\Gamma_L)t_K/2]\,; \end{equation}
$$|2p_D|^2\,W^{Xf}_{D~int}(t_D;t_K)=-|2q_D|^2\,\overline
W^{Xf}_{D~int}(t_D;t_K)
~~~~~~~~~~~~~~~~~~~~~~~~~~~~~~~~~~~~~~~~~~~~~~~~~$$
$$=2|b_{Sf}|^2\,{\rm Re}[a_{-S}^{(X)\ast} a^{(X)}_{+S} \exp(i\Delta
m_D t_D)]\exp[-(\Gamma_+ +\Gamma_-)t_D/2-\Gamma_S t_K]$$
\begin{equation}
~~+2|b_{Lf}|^2\,{\rm Re}[a_{-L}^{(X)\ast} a^{(X)}_{+L} \exp(i\Delta
m_D t_D)]\exp[-(\Gamma_+ +\Gamma_-)t_D/2-\Gamma_L t_K]\,.
\end{equation}
And, at last, there are double-interference contributions, which
contain interference of both $D_\pm$ and $K_{S,L}$:
$$|2p_D|^2\,W^{Xf}_{DK~int}(t_D;t_K)=-|2q_D|^2\,\overline
W^{Xf}_{DK~int}(t_D;t_K)
~~~~~~~~~~~~~~~~~~~~~~~~~~~~~~~~~~~~~~~~~~~$$
$$~~=2\{{\rm Re}[a_{-L}^{(X)\ast} a^{(X)}_{+S} b_{Lf}^\ast
b_{Sf}\exp(i\Delta m_D t_D + i\Delta m_K t_K)]$$ $$~~~+{\rm
Re}[a_{+L}^{(X)\ast} a^{(X)}_{-S} b_{Lf}^\ast b_{Sf}\exp(-i\Delta
m_D t_D + i\Delta m_K t_K)]\}$$ \begin{equation}
~~~\times\exp[-(\Gamma_+ +\Gamma_-)t_D/2 -(\Gamma_S +\Gamma_L)
t_K/2]\,.  \end{equation}
In eqs.(13)-(15) we have used
\begin{equation}
\Delta m_K=m_L-m_S\,,~ \Delta m_D=m_- -m_+\,.
\end{equation}
Of course, we consider $K$-meson decay amplitudes $b_{Sf}$ and
$b_{Lf}$ as known from previous experiments.

Now, if $\Gamma_+\neq\Gamma_-$ and $\Delta m_D\neq0$, eqs.(12)-(15)
show that the double-time distributions for the cascade (1), (8)
contain 10 terms with different time-dependence, which can be
separated from each other. They allow to measure 4 absolute values
of decay amplitudes $a^{(X)}_{\pm S}, a^{(X)}_{\pm L}$ and 6 their
relative phases. To end this section, we express those eigenstate
amplitudes in terms of flavor amplitudes corresponding to flavor
transitions (2), (3):
\begin{equation}
2a^{(X)}_{+S}=\frac{p_D}{p_K}\,a^{(X)}_{DK}+\frac{q_D}{p_K}\,
a^{(X)}_{\overline DK}+\frac{p_D}{q_K}\,a^{(X)}_{D\overline K}
+\frac{q_D}{q_K}\,a^{(X)}_{\overline D\,\overline K}\,  ;
\end{equation}
\begin{equation}
2a^{(X)}_{-L}=\frac{p_D}{p_K}\,a^{(X)}_{DK}-\frac{q_D}{p_K}\,
a^{(X)}_{\overline DK}-\frac{p_D}{q_K}\,a^{(X)}_{D\overline K}
+\frac{q_D}{q_K}\,a^{(X)}_{\overline D\,\overline K}\,  ;
\end{equation}
\begin{equation}
2a^{(X)}_{+L}=\frac{p_D}{p_K}\,a^{(X)}_{DK}+\frac{q_D}{p_K}\,
a^{(X)}_{\overline DK}-\frac{p_D}{q_K}\,a^{(X)}_{D\overline K}
-\frac{q_D}{q_K}\,a^{(X)}_{\overline D\,\overline K}\,  ;
\end{equation}
\begin{equation}
2a^{(X)}_{-S}=\frac{p_D}{p_K}\,a^{(X)}_{DK}-\frac{q_D}{p_K}\,
a^{(X)}_{\overline DK}+\frac{p_D}{q_K}\,a^{(X)}_{D\overline K}
-\frac{q_D}{q_K}\,a^{(X)}_{\overline D\,\overline K}\,  .
\end{equation}
Also useful may be other combinations of amplitudes. For
instance, transitions $D\to K_{S,L}$ and $\overline D\to K_{S,L}$
can be described by the amplitudes
\begin{equation}
a_{DS} = \frac{a_{DK}}{2p_K} + \frac{a_{D\overline K}}{2q_K} =
\frac{a_{+S}+a_{-S}}{2p_D}\,,~~~~
a_{DL} = \frac{a_{DK}}{2p_K} - \frac{a_{D\overline K}}{2q_K} =
\frac{a_{+L}+a_{-L}}{2p_D}\,;
\end{equation}
\begin{equation}
a_{\overline DS} = \frac{a_{\overline DK}}{2p_K} +
\frac{a_{\overline D\overline K}}{2q_K} =
\frac{a_{+S}-a_{-S}}{2q_D}\,, ~~~~
a_{\overline DL} = \frac{a_{\overline DK}}{2p_K} -
\frac{a_{\overline D\overline K}}{2q_K} =
\frac{a_{+L}-a_{-L}}{2q_D}\,. \end{equation}

Single-transition cases simplify the eigenstate amplitudes.
For the pure transition (2)
\begin{equation}
a^{(X)}_{+S}=-a^{(X)}_{-L}\,,~~~
a^{(X)}_{-S}=-a^{(X)}_{+L}\,; ~~~~~a_{DS}=-a_{DL}\,,~~~a_{\overline
DS}=a_{\overline DL}\,;
\end{equation}
while for the pure transition (3)
\begin{equation}
a^{(X)}_{+S}=a^{(X)}_{-L}\,,~~~ a^{(X)}_{-S}=a^{(X)}_{+L}\,;
~~~~~~~~~a_{DS}=a_{DL}\,,~~~a_{\overline DS}=-a_{\overline DL}\,.
\end{equation}
In the general case of two flavor transitions with $CP$-violation,
all four eigenstate (or flavor) amplitudes become independent.

\section {The case of $CP$-conservation}

Now we discuss physical meaning of the obtained expressions in more
detail. At first, for simplicity, we begin with exact conservation
of $CP$-parity which will be assumed throughout the present section,
both for $D$-mesons and for kaons. Then, the eigenstates $K_{S,L}$
have definite $CP$-parities equal to $\pm1$ respectively. $D$-meson
eigenstates should also have definite $CP$-parities, and we suggest
in this section that the indices of $D_\pm$ just label $CP$-parities
$\pm1$ of the eigenstates.

In what follows we need to fix final states for decays (1) and (8).
As the first stage of the cascades we can use decays
\begin{equation}
D^0 (\overline D^0)\rightarrow (\pi^0,\eta,\eta',\rho^0,\omega,
\phi) + K^0 (\overline K^0)\,.
\end{equation}
These final states may be produced by various decay mechanisms but
look similar in terms of the formalism of the preceding section.
Since we assume the exact $CP$-conservation, all the final states
(1) with kaons being in one of their eigenstates have definite
$CP$-parities equal to $(CP)_X(CP)_K(-1)^{S_X}$, where $(CP)_X$ and
$S_X$ are $CP$-parity and spin of the system $X$. For decays (25)
this $CP$-parity is just opposite to the $CP$-parity of the
corresponding kaon eigenstate $(CP)_K$. So, the above choice of
eigenstates $D_\pm$ leads, for all $X$ in (25), to vanishing of
two amplitudes:
\begin{equation} a_{+S}^{(X)}=a_{-L}^{(X)}=0 \end{equation}
(this can be seen also from eqs.(17)-(20)). As a result, all
single-interference contributions (13), (14) disappear. One of
double-interference contributions of eq.(15) disappears as well,
but another survives.

This situation corresponds to existence of two independent decay
branches (instead of four in a general $CP$-violating case)
$$D_+\to X K_L\,,~~ D_-\to X K_S\,;~~~K_{L,S}\to f\,, $$
which can interfere only after the last decay (compare to the
similar consideration in ref.~\cite{az1}). The double-time
distributions, $W^{Xf}(t_D;t_K)$ and $\overline W^{Xf}(t_D;t_K)$
as defined in eq.(11), consist each of two parts, either without
interference of branches or with both $D$ and $K$ interference:
$$4W^{Xf}_{no~int}(t_D;t_K)=4\overline W^{Xf}_{no~int}(t_D;t_K)
~~~~~~~~~~~~~~~~~~~~~~~~~~~~~~~~~~~~~~~~~~~$$
\begin{equation}
~~~~~~~~~~~~~~~~=|a^{(X)}_{+L}\,b_{Lf}|^2 \exp(-\Gamma_+t_D-
\Gamma_Lt_K)+|a^{(X)}_{-S}\,b_{Sf}|^2 \exp(-\Gamma_-t_D-
\Gamma_St_K)\,,
\end{equation}
$$4W^{Xf}_{DK~int}(t_D;t_K)=-4\overline
W^{Xf}_{DK~int}(t_D;t_K)
~~~~~~~~~~~~~~~~~~~~~~~~~~~~~~~~~~~~~$$
$$=2{\rm Re}[a_{+L}^{(X)\ast} a^{(X)}_{-S}\,b_{Lf}^\ast
b_{Sf}\exp(-i\Delta m_D t_D + i\Delta m_K t_K)]\}~~~~~~~~$$
\begin{equation}
\times\exp[-(\Gamma_+
+\Gamma_-)t_D/2 -(\Gamma_S +\Gamma_L)t_K/2]\,.
\end{equation}
One part is monotone (independent contributions of the decay
branches), another oscillates (interference of the branches). These
parts can be easily separated by considering sum or difference of
$W^{Xf}$ and $\overline W^{Xf}$.

The monotone terms determine $D$-meson eigenwidths and absolute
values of amplitudes $|a_{+L}^{(X)}|,\, |a^{(X)}_{-S}|$. Note
that $CP$-conservation makes the $D$-meson indices of amplitudes
and lifetimes be directly and unambiguously related to the
corresponding kaon indices. This means that we can easily determine
$CP$-parity for any $D$-eigenstate. Of course, this $CP$-parity is
just the final state $CP$-parity (we emphasize that it is
single-valued for decays (25) with kaon in an eigenstate, when
$CP$-conservation is exact). The situation is the same as in
ascribing $CP$-parities to $K_S$ and $K_L$ through their decays
to $2\pi$ or $3\pi$. Moreover, this strict correlation between
kaon and $D$-meson indices means that the monotone terms directly
determine relation between an eigenlifetime and $CP$-parity of the
corresponding eigenstate (this problem would not be so simple in
the general case of $CP$-violation; see following sections for
more details).

The oscillating term allows to determine the sign of $\Delta m_D$
in respect to the known sign of $\Delta m_K$. In other words, it
determines which of the $D$-meson eigenstates, $CP$-even or
$CP$-odd, is heavier or lighter. The coefficient of the
oscillating term checks consistency of absolute values of the two
non-vanishing amplitudes, while the constant phaseshift of
oscillations determines the relative phase of these amplitudes.
Note that this double-oscillation (in $t_D$ and $t_K$) is, in
essence, similar to secondary oscillations in kaon regeneration
(ref.~\cite{KO}) which opened possibility to determine the sign of
$\Delta m_K$ in respect to the sign of the regeneration phase.  We
emphasize that the oscillating term allows to relate the heavier
or lighter mass to eigenstate $CP$-parities, but not directly to
longer or shorter lifetimes.

The above expressions have seemingly the same form as for
$B$-meson decays~\cite{az1} where the final strangeness is
strictly correlated with the initial flavor. The real difference
is the independence of amplitudes $a^{(X)}_{-S}$ and $a^{(X)}_{+L}$
if both transitions, (2) and (3), are present. As a result, the two
nonvanishing flavor transitions lead to $|a^{(X)}_{+L}|
\neq|a^{(X)}_{-S}|\,$.

There is one more consequence: complexities of the amplitudes
$a^{(X)}_{+L}$ and $a^{(X)}_{-S}$ (or, equivalently, of the
amplitudes $a^{(X)}_{DK}$ and $a^{(X)}_{D\overline K}$) may be,
generally, different.  This statement looks rather evident for the
final states $\pi^0 K^0(\overline K^0)$ or $\rho^0 K^0(\overline
K^0)$ which combine two isotopic-spin states. It is less familiar
but also true for such final states as, say, $\omega K^0(\overline
K^0)$ which are different components of the same isotopic-spin
state. The reason is that the standard idea of the decay amplitude
having the same phase as the elastic scattering amplitude for
hadrons in the decay final state is not always correct. It is true
only if the final-state interaction (FSI) cannot rescatter the
particular state into some other states. However, the $D$-meson
mass is high enough, and any particular final state in $D$-decay
does can rescatter (most evidently, $K \omega$ may rescatter into
$K\,3\pi$ with pions out of resonance).

Formally, this means that any particular final state in decays (25)
does not diagonalize the strong-interaction $S$-matrix and does not
produce a universal FSI-phase. The amplitudes $a^{(X)}_{DK}$ and
$a^{(X)}_{D\overline K}$ appear to be some linear combinations of
amplitudes for transitions into combined states which diagonalize
the strong $S$-matrix (and produce universal FSI-phases). Since the
mechanisms of transitions (2) and (3) are different, the two
combinations are also different and, therefore, final-state
interactions may generate different phases for the two amplitudes.
Thus, the factor $a^{(X)\ast}_{+L} a^{(X)}_{-S}$ in eq.(28) is real
if only one of the transitions, (2) or (3), is operative, but may
be, generally, complex if the both transitions are possible.
Experimental determination of such phase difference could be useful
to reveal possible decay mechanisms.

Let us discuss specific final states for the second stage (8) of
cascades. If we take $f=2\pi$, then $b_{L(2\pi)}=0$ and the
double-time distributions (11) contain only one term, expressed
through $|a^{(X)}_{-S}|^2$. For $f=3\pi$ we have $b_{S(3\pi)}=0$.
The distributions, again, contain only one term, expressed through
$|a^{(X)}_{+L}|^2$. These two final states lead just to the
situations discussed in ref.~\cite{by}. Cascades with the two final
states allow one to measure absolute values of the corresponding
first-stage amplitudes, but not their relative phase. Therefore,
they do not allow to find amplitudes $a^{(X)}_{DK}$ and $a^{(X)}_{D
\overline K}$ unambiguously.

Semileptonic kaon decays with $f=\pi^\mp l^\pm \nu(\overline\nu)$
have $|b_{Lf}|=|b_{Sf}|$, and all three terms of eqs.(27), (28)
appear in the double-time distributions. Generally, they have
different dependence on $t_D$ which might help to separate them.
In any case, the three terms have different dependence on $t_K$ and
may be separated for sure. Then, one can determine here not only
absolute values of $a^{(X)}_{+L}$ and $a^{(X)}_{-S}$, but their
relative phase as well. In other words, cascades with semileptonic
secondary decays allow one to unambiguously find both the absolute
values and relative phase of the amplitudes $a^{(X)}_{DK}$ and
$a^{(X)}_{D\overline K}$ for the primary decays (25).

Thus, investigation of double-time distributions in cascade decays
(25), (8) may solve several important problems: it measures the $D
$-meson eigenwidths $\Gamma_\pm$ and mass difference $\Delta m_D$,
relates eigenwidths and eigenmasses to each other and to $CP
$-parities of eigenstates, determines amplitudes $a^{(X)}_{D
\overline K}$ and $a^{(X)}_{DK}$ (together with their relative
phase) for the favored and suppressed flavor transitions (2), (3).

Earlier, in refs.~\cite{az1,ars}, we noticed that interesting
problems for $B$-mesons may be attacked also in single-time
distributions over $t_K$ (integrated over $t_B$). Presence of two
transitions in $D$-meson decays produces more of independent
amplitudes and makes single-time distributions less efficient.
Consider, e.g., contribution (28). After integration over $t_D$ it
contains the factor
\begin{equation}
\cos(\phi^{(Xf)}_{SL}-\alpha_D+\Delta m_K t_K)\,,
\end{equation}
where $\phi^{(Xf)}_{SL}$ is the relative phase of
$a^{(X)}_{-S} b_{Sf}$ and $a^{(X)}_{+L} b_{Lf}$, while
$$\tan(\alpha_D/2)=x_D\equiv\frac{2\Delta m_D}{\Gamma_++\Gamma_-
}\,.$$ Single-time distributions in $D$-meson decays provide no
way to separate $\phi^{(Xf)}_{SL}$ from $\alpha_D$. Note, for
comparison, that the $B$-meson analog of $\phi^{(Xf)}_{SL}$ has
a definite value depending on spin and $CP$-parity of the system
$X$ and on the final state $f$ in the secondary kaon decay (it is
$0$ or $\pi$ for semileptonic kaon decays). This is the reason why
$\alpha_D$ cannot be measured model-independently in single-time
decay distributions for neutral $D$-mesons, while similar
distributions in decays of neutral $B$-mesons may be sufficient to
measure an analogous quantity $\alpha_B$~\cite{ars}.

If $\Delta m_D$ and $\Delta\Gamma=\Gamma_+-\Gamma_-$ are vanishing
(or too small to be measured) then the three terms in contributions
(27), (28) have the same $t_D$-dependence. Neutral $D$-mesons in
this situation are unmixed, and so, their decays exactly
correspond to such decays of charged $D$-mesons as, e.g.,
\begin{equation} D^\pm\to(\pi^\pm,\rho^\pm)+ K^0(\overline K^0)
\end{equation}
with subsequent semileptonic kaon decays. The three terms in time
distributions can still be separated by their different
$t_K$-dependence. Note that the single-time distribution in
$t_K$ is sufficient here to separate and measure amplitudes of
transitions (2), (3) and their relative phase. Thus, for unmixed
$D$-mesons the secondary-decay distribution appears to be even more
interesting than the primary-decay one. For the above measurements
one does not need to study the large-$t_K$ region ($t_K\gsim\tau_L
$). Necessary is only the interval of $t_K$ up to about $(10-15)
\tau_S$, overlapping the $K_{S,L}$ interference region.

\section{$CP$-parity eigenstates with $CP$-violation}

To consider the general case which corresponds to violated
$CP$-parity we return to exact expressions (12)-(15) for
double-time distributions. They contain 10 different terms which,
in principle, determine absolute values of 4 amplitudes $a^{(X)}_{
\pm S}, a^{(X)}_{\pm L}$ and 6 their relative phases. To interpret
results of measurements, some physical identification of eigenstates
appears to be necessary. For clarification of this point let us
compare kaons and heavier mesons.

First of all, note that every meson eigenstate has 3 main
characteristics: width, mass and $CP$-parity (at least,
approximate). Thus, we have 3 different ways of labeling two
eigenstates by 3 corresponding pairs: shorter or longer lifetime;
lighter or heavier mass; even or odd $CP$-parity. Of course, these
ways are physically equivalent, but the equivalence can be realized
only if one has experimental methods to relate those 3
characteristics with each other.

The kaon eigenstates, $K_{S,L}$, are usually identified and labeled
by their lifetimes, shorter or longer. Their prevailing hadronic
decay modes, $2\pi$ or $3\pi$, determine their $CP$-parity, at
least approximately. The mass difference of $K_S$ and $K_L$ can be
easily measured in semileptonic decays. On this way, however, one
cannot find the sign of $\Delta m_K$, i.e., to determine which of
the states is heavier or lighter. Only the specially invented (and
rather complicated) experiments allowed to measure the sign of mass
difference $\Delta m_K$ and, thus, relate widths and masses of
$K_S$ and $K_L$ to each other (ref.~\cite{KO}; more detailed
theoretical references and compilation of experimental results see
in ref.~\cite{PDTold}). This provided possibility for unambiguous
measuring the kaon $CP$-violating parameters (note that the signs
of $\phi_{+-}$ and $\phi_{00}$, the phases of $\eta_{+-}$ and
$\eta_{00}$, are measured only in respect to the sign of $\Delta
m_K$). As a result, we have now, indeed, at least three equivalent
ways of identifying $K_{S,L}$: by shorter or longer lifetime, by
heavier or lighter mass, by the (approximate) $CP$-parity.

The first of these ways cannot be applied at present to $D$-meson
eigenstates, $D_\pm$, because of very small (and yet unobserved)
difference of eigenlifetimes (the same is true for $B$-mesons). The
absolute value of $\Delta m$ has been measured for $B_d$-mesons
(results and references see in ref.~\cite{pdt}), there are
suggestions how to do this for $D$-mesons as well. In contrast to
kaons, for $B$- and $D$-mesons the values of $|\Delta m|$ are
expected to be noticeably higher than $|\Delta\Gamma|$ (it is
definitely so for $B$-mesons). Therefore, a rather familiar way in
the current literature is to identify eigenstates, for both $B$ and
$D$, as heavier and lighter (i.e., $B_{h,l}$ and $D_{h,l}$). On the
other side, identification of eigenstates by their approximate $CP
$-parities was suggested in ref.~\cite{az1} for $B$-mesons and may
be applied to $D$-mesons as well. The real problem is how to relate
with each other the different approaches to the $B$- and/or $D
$-meson eigenstates. For $B$-mesons it was discussed in
ref.~\cite{az3}. Here we consider the situation for $D$-mesons and
the role of their cascade decays.

For definiteness, we use at the first stages of cascades the same
decays (25) as in the preceding section.  At the second stages we
may also use, as before, the three typical kinds of kaon decays:
either semileptonic, or purely pionic with 2 or 3 pions produced.
As we have seen, only semileptonic kaon decays could allow to
measure the relative phase of amplitudes for $CP$ conserved. On the
contrary, with violated $CP$ we might, principally, use any of the
three decay modes, since for all of them \mbox{$|b_{Sf}|,|b_{Lf}|
\neq0$}. However, decays $K^0(\overline K^0)\to3\pi$ are still
really useless because of too small $|b_{S(3\pi)}|$. When comparing
decays $K^0(\overline K^0) \to2\pi$ to semileptonic ones, the
semileptonic decays may appear experimentally more favorable,
by the same arguments as suggested in $B$-meson
studies~\cite{ars,bk,az3}. This problem, however, will not be
discussed here anymore since it requires detailed investigation
for a particular detector.

To discuss possible measuring procedures we begin with a
hypothetical suggestion that eigenwidths $\Gamma_+$ and $\Gamma_-$
are different enough, so that every term in expressions (12)-(15)
can be extracted and studied separately. We also assume that all
$K$-meson parameters and decay amplitudes are known. Then, first of
all, from monotone terms of eq.(12) we find two eigenwidths $
\Gamma_\pm$ and four absolute values of amplitudes $| a^{(X)}_{\pm
S}|, |a^{(X)}_{\pm L}|$. Their kaon indices $S,L$ are fixed by the
corresponding exponentials in $t_K$. However, this is not so for
$D$-meson indices $\pm$, which (contrary to the $CP$-conservation
case) are not unambiguously related with kaon ones and not
determined by $t_K$-dependence.

Now we can specify possible meaning of the indices $\pm$, which
has not been fixed yet, and define how to ascribe them to
amplitudes and eigenwidths. If $CP$-violation is small indeed (or
at least effectively), we may fix the indices as showing
approximate $CP$-parities of eigenstates. Namely, in such a case
there should be two larger and two smaller amplitudes, and the
indices $\pm$ of the decaying eigenstates may be ascribed (for
states $X$ in decays (25)) so that larger amplitudes conserve
$CP$-parity:
\begin{equation}
|a^{(X)}_{-S}| > |a^{(X)}_{+S}|\,,~~~~~~~
|a^{(X)}_{+L}| > |a^{(X)}_{-L}|\,.  \end{equation}
Note that in presence of only one transition, (2) or (3), we have
$|a^{(X)}_{\pm S}|=|a^{(X)}_{\mp L}|$ (see eqs.(23),(24)), and only
one of the inequalities is independent.

At first sight, the two inequalities (31) look trivial even for a
general case, since in every pair of amplitudes one of their
absolute values is, as a rule, greater than another. However, an
essential and nontrivial property of the inequalities is that the
two larger amplitudes must correspond to different eigenstates of
both kaons and $D$-mesons (in the expression (12) their monotone
contributions should contain exponentials in $t_D$ and $t_K$ with
"opposite" combinations of $D$-meson and kaon eigenwidths in the
exponents; this should and may be checked). Really, one pair of
amplitudes with the same kaon index (e.g., $S$) would be sufficient
to ascribe indices $\pm$ to the $D$-meson states. Then the
corresponding $t_D $-exponentials determine, which of $D$-meson
eigenwidths is $\Gamma_+$ and which is $\Gamma_-$; in other words,
this procedure relates eigenwidths and approximate $CP$-parities of
the eigenstates. After that the indices for another pair of
amplitudes are completely fixed, and the second inequality (31) may
appear true or false. In the case of small $CP$-violation it should
be true, of course.

If, however, the inequalities are inconsistent, then the choice
(31) is contradictory. In such a case the $CP$-violation in
transitions $D_\pm\to K_{S,L}$ could not be considered as
effectively small (similar problems for $B$-mesons are discussed
in refs.~\cite{az1,ars,az3}). The approximate $CP$-parities of the
eigenstates $D_\pm$ would become mode-dependent, i.e. the effective
$CP$-parity for the same eigenstate would be different when
determined from transitions to $K_S$ or $K_L$ (or some other final
states with definite $CP$-parities). Similar situation is well
known for the space-parity violation in weak interactions (recall,
that the kaon parity is mode-dependent: it is different when
determined from decays $K\to2\pi$ or $K\to3\pi$).

Now, let us stick to a definite prescription of $CP$-parities based
on a particular decay mode. As the next step we may use two terms
of expression (13) to find unambiguously the phases
arg$(a^{(X)\ast}_{+L}a^{(X)}_{+S})$ and
arg$(a^{(X)\ast}_{-L}a^{(X)}_{-S})$. Signs of these phases are
determined in respect to the known sign of $\Delta m_K$. Note that
if only one of transitions, (2) or (3), is operative then the two
phases differ only by the sign.

If we use two terms of expression (14) to find arg$(a^{(X)\ast
}_{-S} a^{(X)}_{+S})$ and arg$(a^{(X)\ast}_{-L} a^{(X)}_{+L})$
we discover that their signs could be measured only in respect
to the yet unknown sign of $\Delta m_D$. If, again, only one of
transitions, (2) or (3), worked, the situation would become
definite due to equality of ratios
$$\frac{a^{(X)\ast}_{-S}a^{(X)}_{+S}}{a^{(X)\ast}_{+L}
a^{(X)}_{+S}}= \frac{a^{(X)\ast}_{-L}a^{(X)}_{+L}}{a^{(X)\ast}_{-L}
a^{(X)}_{-S}}$$ (which would equal to -1 or +1 respectively for
transitions (2) or (3); see relations (23) and (24)).
As a result, the relative phase of two terms in expression (14)
equals to the relative phase of two terms in expression (13). This
phase may be measured from $t_K$-dependence of the contribution
(13), the sign of the phase being determined in respect to $\Delta
m_K$. Then, $t_D$-dependence of the contribution (14) determines
the sign of $\Delta m_D$ in respect to the sign of the (now known)
phase, i.e. really in respect to the sign of $\Delta m_K$. Such
determination can be achieved also when both transitions are
present, but not so easily, since then the expressions (13) and
(14) contain different phases. Note, however, that the
contributions (13), (14) are suppressed if $CP$-violation is small
in any sense.

Contributions (15) are, even by themselves, sufficient to find
the sign of $\Delta m_D$ directly in respect to the sign of $\Delta
m_K$.  Indeed, inequalities (31) lead to \begin{equation}
|a^{(X)\ast}_{+L}\,a^{(X)}_{-S}| > |a^{(X)\ast}_{-L}\,a^{(X)}_{+S}|
\end{equation}
and allow to discriminate the two terms in (15); after that the
double-time dependence directly compares $\Delta m_D$ to $\Delta
m_K$ and, in particular, determines their relative sign. Phases
arg$(a^{(X)\ast}_{-L}\,a^{(X)}_{+S})$ and
arg$(a^{(X)\ast}_{+L}\,a^{(X)}_{-S})$ can be also determined here.
They check self-consistency of the procedure since, surely, there
should be $${\rm arg}(a^{(X)\ast}_{+L}\,a^{(X)}_{+S})- {\rm
arg}(a^{(X)\ast}_{-S}\,a^{(X)}_{+S})= {\rm
arg}(a^{(X)\ast}_{+L}\,a^{(X)}_{-S})\,,$$ \begin{equation} {\rm
arg}(a^{(X)\ast}_{-L}\,a^{(X)}_{-S})+ {\rm
arg}(a^{(X)\ast}_{-S}\,a^{(X)}_{+S})= {\rm
arg}(a^{(X)\ast}_{-L}\,a^{(X)}_{+S})\,.  \end{equation}
Even if the choice (31) is contradictory, we still may define
eigenstate $CP$-parities (i.e., ascribe the indices $\pm\,$) so to
provide the inequality (32).

The procedures described remind what was really done in kaon
studies. For each eigenstate they allow to relate together various
state's properties: shorter or longer lifetime, positive or
negative (approximate) $CP$-parity, and heavier or lighter mass.
Of course, their combination could be fixed also by different
(though equivalent) procedures. We emphasize, however, that some
physical procedures are necessary and inevitable. Only with such
procedures one becomes able to measure flavor-transition
amplitudes unambiguously. We will see further in this section that
the same is true also for $CP$-violating parameters.

The physical necessity of $CP$-parity prescriptions for eigenstates
may be traced to the following simple reason. Time dependence
(single or double) is always related to eigenstates. On the other
side, flavor amplitudes (say, $a^{(X)}_{DS}, a^{(X)}_{\overline
DS}$) correspond to flavor states $D$ and $\overline D$, which
are linear combinations  of eigenstates. $D$ is conventionally
considered as $$D\sim D_++D_-\,.$$ This definition of $D$ is
insensitive to accurate identification of eigenstates.  $\overline
D$, contrary, is proportional to the difference of the eigenstates,
$$\overline D\sim D_+-D_-\,,$$ and their interchange would change
the sign of $\overline D$. To cope with the conventional relation
$\overline D=CP(D)$ we should apply some procedure to define
$CP$-parities of eigenstates, and then subtract the $CP$-odd state
from the $CP$-even one. Without any procedure the state $\overline
D$, and various related physical quantities as well, can be
determined only up to the sign.

Let us discuss now a more realistic situation when $\Gamma_+ =
\Gamma_-$ with available precision. In such a case the four
amplitudes $a^{(X)}_{\pm S}, a^{(X)}_{\pm L}$ cannot be determined
unambiguously since several contributions have the same
$t_D$-dependence (see, e.g. eqs.(12), (13)) and cannot be
completely separated. The $t_D$-dependence becomes the same for
every contribution if $\Delta m_D$ is also too small and physical
discrimination of eigenstates $D_\pm$ disappears at all. However,
$t_K$-dependencies of different contributions are still different,
and partial separation of various contributions is still possible.
Indeed, by comparing decays of initially pure states $D^0$ and
$\overline D^0$ one could separate contributions (12), (13) on one
side and (14), (15) on the other. Then, by means of different
$t_K$-dependence we could discriminate (13) from (12) and split
(12) into two parts. In the same manner (15) would be discriminated
from (14), which is also split into two parts. So, after all, we
can split decay time-distributions for $D^0$ and $\overline D^0$
only to 6 different terms (instead of 10 for $\Delta\Gamma_D\neq0,
\Delta m_D\neq0$). One may be still able to find amplitudes of
transitions (2), (3), but only with additional simplifying
assumptions (e.g., neglecting $CP$-violation or describing it by
some special models).

To understand the situation we return to the cascade amplitudes of
eqs.(9), (10). If $\Delta\Gamma_D=\Delta m_D=0$, then
$e_-(t_D)=e_+(t_D)$; mixing is absent, and the initial $D$-meson
state decays without evolution. Therefore, more adequate are not
the eigenstate amplitudes $a^{(X)}_{\pm L},\,a^{(X)}_{\pm S}$, but
their combinations $a^{(X)}_{DS}$ and $a^{(X)}_{DL}$ determined by
eqs.(21) (or $a^{(X)}_{\overline DS}$ and $a^{(X)}_{\overline DL}$;
see eqs.(22)) which correspond to transitions $D^0\to K_{S,L}$ and
$\overline D^0\to K_{S,L}$ without mixing of initial $D^0$ and/or
$\overline D^0$. Amplitudes of every pair are still coherent to
each other, but not coherent to amplitudes of another pair.
Therefore, instead of 10 physical quantities we have now only 6
measurable quantities, which are two absolute values and one
relative phase in each of the amplitude pairs (21), (22). Thus
again, the 6 physically meaningful quantities correspond to 6
separable terms in double-time distributions (12)-(15) at $\Delta
\Gamma_D=\Delta m_D=0$ (when returning to the case of
$CP$-conservation, we would have additional relations $a^{(X)}_{DS}
=-a^{(X)}_{\overline DS},\,a^{(X)}_{DL}= a^{(X)}_{\overline DL}$
for any $X$ in decays (25), further diminishing the number of
independent physical quantities).

Note one more specific feature of cascade decays (1), (8). Each of
them contains two $CP$-violating parameters which may be
phenomenologically independent. Physically, they correspond to
$CP$-violation in transitions (2) and (3). Even rough
estimates~\cite{ai} show that $CP$-violation in the suppressed
transition (3) may appear greater than in the favored transition
(2). It may also be very sensitive to some New Physics.

As phenomenological $CP$-violating parameters for cascades (25),
(8) one can use, e.g., the ratios \begin{equation}
\eta^{(X)}_{DS}=\frac{a^{(X)}_{+S}}{a^{(X)}_{-S}}\,,~~~~~
\eta^{(X)}_{DL}=\frac{a^{(X)}_{-L}}{a^{(X)}_{+L}}
\end{equation}
vanishing in the limit of exact $CP$-conservation (for both
$D$-meson and kaon decays). They have the same structure as the
standard parameters $\eta_{00}$ and $\eta_{+-}$ for neutral kaon
decays: each of them is the ratio of two amplitudes,
$CP$-suppressed and $CP$-favored, with the same final state and
different decaying eigenstates. If $CP$-parities of $D$-meson
eigenstates can be chosen so to satisfy inequalities (31) then,
for the final states (25), both parameters (34) have absolute
values smaller than unity. Inconsistency of the inequalities (31)
would imply that the absolute value is less than unity for one of
the parameters, but greater than unity for another. In any case,
$CP$-parities of eigenstate can be chosen so to satisfy the
condition (32). In terms of the $\eta$-parameters this condition
takes the simple form $$|\eta^{(X)}_{DS}\,\eta^{(X)}_{DL}|<1\,.$$
Though this inequality looks quite natural, we emphasize that
generally it may be nontrivial. Its correctness for a particular
system $X$ in decay (1) may always be achieved  by a special choice
of $CP$-properties of $D$-meson eigenstates in this particular
decay. Note, however, that such special choice might depend on
the system $X$ if inequalities (31) are inconsistent (recall,
that inconsistency of (31) would imply mode-dependence of
$CP$-properties for $D$-meson eigenstates).

One more possible way to describe $CP$-violation, appropriate for
any cascade (1), is to use parameters of the kind
\begin{equation}
\lambda^{(X)}_{DS}=\frac{q_D}{p_D}\,\frac{a^{(X)}_{\overline DS}}
{a^{(X)}_{DS}}\,,~~~~~ \lambda^{(X)}_{DL}=\frac{q_D}{p_D}\,
\frac{a^{(X)}_{\overline DL}}{a^{(X)}_{DL}}\,.  \end{equation}
$CP$-violation may be measured by their deviation from
$CP$-conserving values, which are $+1$ or $-1$. For the
final states (25) the $CP$-conserving values are clearly seen
from the relations
$$\lambda^{(X)}_{DS}=\frac{\eta^{(X)}_{DS}-1}{\eta^{(X)}_{DS}+1}
\,,~~~~~\lambda^{(X)}_{DL}=\frac{1-\eta^{(X)}_{DL}}{1+\eta^{(X)
}_{DL}}\,.$$ By using relations (17)-(22) one can easily express
the parameters (35) through flavor-transition amplitudes. In the
case of a single-flavor transition, (2) or (3), the two parameters
$\lambda^{(X)}_{DS}$ and $\lambda^{(X)}_{DL}$ differ only in sign;
they are proportional to the ratio of the corresponding
flavor-transition amplitudes for $\overline D$ and $D$. In the
presence of both transitions (2) and (3) we still can describe
their $CP$-properties by separate parameters $$\lambda^{(X)}_{D
\overline K}=\frac{q_D}{p_D}\,\frac{q_K}{p_K}\, \frac{a_{\overline
DK}}{a_{D\overline K}}\,,~~~~~~
\lambda^{(X)}_{DK}=\frac{q_D}{p_D}\,\frac{p_K}{q_K}\,
\frac{a_{\overline D\overline K}}{a_{DK}}$$
for each transition (just such parameters were used in our papers
on $B$-mesons~\cite{az1,ars,ad,az2,az3}). If $CP$-violation in the
two transitions is the same (i.e. $\lambda^{(X)}_{D\overline K}=
\lambda^{(X)}_{DK}$), then, again, we have $\lambda^{(X)}_{DS}=
-\lambda^{(X)}_{DL}$.

In difference with the $CP$-violating parameters (34), parameters
(35) do not contain explicitly $D$-meson eigenstates. However, they
use the states $D$ and $\overline D$. Therefore, as explained above,
they also cannot be determined unambiguously without some
$CP$-prescription for $D$-meson eigenstates.

Such property is not unique for decays into neutral kaons.
Consider, e.g., decays
\begin{equation}
D^0(\overline D^0)\to F
\end{equation}
with amplitudes $a^{(F)}_D,\, a^{(F)}_{\overline D}$; here $F$ is
some final state with a definite $CP$-parity \footnote{Really we
briefly repeat here a similar discussion of ref.\cite{ars} for
$B$-mesons.}. The time distributions of decays (36) contain terms
proportional to
\begin{equation}
{\rm Re}\lambda^{(F)}_{D}\,\sinh[(\Gamma_+-\Gamma_-)t]~~~~~~~~
{\rm and}~~~~~~~~
{\rm Im}\lambda^{(F)}_{D}\,\sin[(m_+-m_-)t]\,,
\end{equation}
where
\begin{equation}
\lambda^{(F)}_D=\frac{q_D}{p_D}\,\frac{a^{(F)}_{\overline
D}}{a^{(F)}_D}\,.  \end{equation} Expressions (37) show that the
sign of Re$\lambda$ can be determined experimentally only if we
know relation between eigenwidths and (approximate) $CP$-parities
of eigenstates, while the sign of Im$\lambda$ needs relation
between eigenmasses and $CP$-parities. An essential point is that
in practice we cannot find these relations in decays (36)
themselves (especially for masses), while cascade decays (1), (8)
with intermediate neutral kaons may provide such possibilities.

\section{Mass eigenstates with $CP$-violation}

In preceding sections we have demonstrated that identification of
$D$-eigenstates by their $CP$-parities, exact or approximate,
requires two procedures for complete description of the states. One
of them uses monotone (in $t_D$ and $t_K$) terms of decay
distributions and relates eigenwidths to the corresponding
$CP$-parity eigenstates. Another procedure, by means of double
oscillations (again, in $t_D$ and $t_K$), relates eigenmasses to
the eigenstates and determines which of them is heavier or lighter.

However, we have mentioned above that a rather familiar approach
in the current literature is to identify eigenstates of heavy
flavored neutral mesons from the beginning by their masses, as
heavier (e.g., $D_h$) or lighter (e.g., $D_l$) states. In such
notations the kaon nomenclature would look as $K_h,\,K_l$ instead
of $K_L,\,K_S $ correspondingly. Evidently, this approach is
meaningful only if $\Delta m$ can be measured, even if $\Delta
\Gamma$ is too small for measuring. In this section we will
consider $D$-meson eigenstates as identified by their masses.

Formally, cascade amplitudes and double-time distributions with
such identification of eigenstates can be easily obtained from eqs.
(9), (10) and (12)-(15), respectively. Of course, meaning of the
indices $\,\pm$ should be defined differently than in the preceding
section. If we take, by definition, $\Delta m_D>0$, then according
to eq.(16) we need to identify the states as
\begin{equation} D_-
\equiv D_h\,,~~~~D_+\equiv D_l \end{equation}
(similar to kaons). Therefore, we should rewrite eigenstate
amplitudes as \begin{equation}
a^{(X)}_{-L}\to a^{(X)}_{hL}\,,~~~~ a^{(X)}_{-S}\to
a^{(X)}_{hS}\,,~~~~ a^{(X)}_{+L}\to a^{(X)}_{lL}\,,~~~~
a^{(X)}_{+S}\to a^{(X)}_{lS}\,. \end{equation} Here the first
subscripts correspond to initial $D$-meson eigenstates being
heavier or lighter, while the second ones are for final kaon
eigenstates, Long- or Shortliving.

At first sight, the above change being supplemented by the
substitution $$\Gamma_- \to \Gamma_h\,,~~~~~ \Gamma_+ \to
\Gamma_l $$ should be quite sufficient. The situation, however, is
not so simple. The real problem, as before, is how to construct
procedures that determine, which of phenomenological amplitudes is
which, and allow to relate physical eigenwidths and $CP$-properties
with the mass labels of eigenstates.

To discuss this problem we, again, begin with a hypothetical
assumption that every particular contribution in double-time
distributions (12)-(15) can be separated experimentally by detailed
study of double-time distributions for initial $D$- and $\overline
D$-states.  This assumes also that both $\Delta m_D$ and
$\Delta\Gamma_D$ are large enough to be measurable.

First of all, we separate terms having monotone behavior in $t_K$
(see eqs.(12), (14)). They correspond to $D$-meson decays with
production of $K_S$ or $K_L$, without their interference. Both
cases lead to $t_D$-dependence of the form
$$|a_1|^2\,\exp(-\Gamma_1t_D)+
|a_2|^2\,\exp(-\Gamma_2t_D)~~~~~~~~~~~~~~$$
\begin{equation}
~~~~~~~~+2|a_1a_2|\cos(\phi_{12} +\Delta
m_Dt_D)\exp[-(\Gamma_1+\Gamma_2)t_D/2]\,.
\end{equation}
Here $a_1$ and $a_2$ are two amplitudes (e.g., $a^{(X)}_{hL},\,
a^{(X)}_{lL}$ for terms with the factor $\exp(-\Gamma_L t_K)$ in
eqs.(12), (14)), $\phi_{12}$ is their relative phase; $\Gamma_1,\,
\Gamma_2$ stay for $\Gamma_h$ and/or $\Gamma_l$.

We see that the distribution (41), even having been ideally
measured, would allow to determine the two amplitudes $a_1$ and
$a_2$ (up to complex conjugation, because of possible change $a_1
\to a_2^\ast,\, \Gamma_1\to\Gamma_2$) and relate them with
eigenwidths $\Gamma_1, \,\Gamma_2$, but could not show which of
them is for heavier or lighter $D$-meson eigenstates. A formal
reason is that the contributions (12) and (14) have  symmetry
properties: they do not change under substitution, e.g., $a^{(X)}_{
+L}\to a^{(X)\ast}_{-L},\, \Gamma_+\to\Gamma_-$ (and/or similar for
subscript $S$), with $\Delta m_D$ staying unchanged. Thus,
interchange of states $D_+$ and $D_-$ (i.e. of states $D_l$ and
$D_h$) without changing $\Delta m_D$ could not have any influence
on contributions (12) and (13).

So, the $t_K$-monotone contributions in decay time distributions
cannot discriminate the two eigenstates, heavier and lighter: they
cannot relate the measured widths $\Gamma_1$ and $\Gamma_2$ to
heavier or lighter eigenstates and cannot relate amplitudes to the
corresponding eigenstate transitions. Of course, study of such
terms would not be, nevertheless, useless: it can determine
absolute values of all four amplitudes and two of their relative
phases (one may note, however, that the phases can be determined
in this way only up to signs).

Contributions (13), oscillating in $t_K$ without oscillations in
$t_D$, determine more of relative phases, but cannot yet, by
themselves, distinguish amplitudes for initial states $D_h$ or
$D_l$. The reason is that these contributions also satisfy a
symmetry property preventing, again, discrimination of $D_\pm$
(i.e. of $D_h, D_l$): they do not change under substitution $a^{(X)
}_{+L}\to a^{(X)}_{-L},\, a^{(X)}_{+S}\to a^{(X)}_{-S},\,\Gamma_+
\to\Gamma_-$. This symmetry is different from one discussed above.
It works, nevertheless, for contributions (12) as well, but not for
contributions (14). Therefore, contributions (13) together with
(12) and (14) could, in principle, distinguish the states $D_h$ and
$D_l$ and relate them to amplitudes and eigenwidths. Note, however,
that the contributions (13), (14) vanish in the case of
$CP$-conservation and, thus, are expected to be small (this brief
discussion may be directly compared with a similar discussion in
the preceding section).

The situation can be really resolved by contributions (15),
oscillating in both $t_K$ and $t_D$. These contributions by
themselves violate the both above symmetries. Here the interchange
of amplitudes and widths for $h$- and $l$-states would require
simultaneous change of the relative sign between the known terms
proportional to $\Delta m_D$ and $\Delta m_K$. Therefore, here at
last we can determine which two of four amplitudes correspond to
decays of, say, $D_h$. After that, all the relative phases of the
four amplitudes become unambiguous as well. Eigenwidths have been
earlier related to definite amplitudes which now become specified
as $h$- and $l$-amplitudes. So, due to contributions (15) the
widths can be definitely related to $h$- and $l$-states. The
$CP$-parity (exact or approximate) of eigenstates becomes also
determined just by contributions (15), through relation between
$|a^{(X)}_{lL}a^{(X)}_{hS}|$ and $|a^{(X)}_{lS}a^{(X)}_{hL}|$
which should correspond to inequality (32) (without substitution
(40)).

When discussing $CP$-violation in terms of $D_h, D_l$ the
parameters (34) are not convenient, since without any special model
one does not know {\it a priori} which of amplitudes, e.g.
$a^{(X)}_{hS}$ or $a^{(X)}_{lS}$, is larger in absolute value.
In contrast, the parameters (35) are quite appropriate. One should
note, however, that relations (17)-(22) between flavor and
eigenstate amplitudes were written for the identification
of $D$-eigenstates by their $CP$-parity (at least, in one pair of
transitions $D_\pm\to K_S$ or $D_\pm\to K_L$). The substitutions
(39), (40) should not be applied to them directly, before the
amplitudes having been identified as described above in this
section. Therefore, without using double-flavor oscillations one
cannot unambiguously extract flavor-transition amplitudes from
distributions (12)-(15) and, as a result, cannot unambiguously
determine parameters (35) (and (38) as well). Similar discussions
for $B$-mesons see in refs.~\cite{ars,az2,bk,az3}.

We emphasize once more that there is no direct way to relate
eigenwidths (even if measured) to a heavier (or lighter)
eigenmass. The widths may be directly related only to amplitudes.
The role of the coherent double-flavor oscillations is to relate
the amplitudes (and, hence, the widths as well) to the mass
eigenstates. The same note is true for $CP$-properties of the
eigenstates.  $CP$-violating parameters cannot be determined
unambiguously without using the double-flavor oscillations.

Let us briefly discuss the problem of flavor amplitudes for
transitions (2), (3) with mass labeling of $D$-meson eigenstates.
Formally, they can be easily expressed through eigenstate
amplitudes by eqs. (17)-(22). However, $D$-meson subscripts $\pm$
in these relations correspond just to $CP$-properties of
eigenstates and should not be changed according to substitution
(39). Therefore, the flavor amplitudes can be determined
unambiguously only when (approximate) $CP$-parities of the mass
eigenstates have been measured, as can be done in double-flavor
oscillations. The reason is the same as discussed in the preceding
section: to construct correct flavor states ($\overline D$ in the
conventional approach) we need to know which of states, $D_h$ or $D_l$,
is (approximately) $CP$-even  and/or $CP$-odd.

\section{Concluding remarks. Strategy of measurements}

In previous sections we have shown that coherent double-flavor
oscillations suggest possibilities to solve various problems in
$D$-meson physics.  To understand which experiment may study this
or that problem, we begin this section with estimating expected
values of different effects.

The Minimal Standard Model leads to a natural estimate
$$\left|\,\frac{a^{(X)}_{DK}}{a^{(X)}_{D\overline K}}\,\right|
\approx\left|\,\frac{a^{(X)}_{\overline D\overline K}}{a^{(X)}_{
\overline DK}}\,\right|\sim {\cal O}(\tan^2\theta_C)\,,$$ with
$\tan^2\theta_C\approx0.05$, where $\theta_C$ is the Cabibbo angle.
This expectation corresponds to known experimental data~\cite{pdt}
and, due to relations (17)-(22), leads to boundaries $$|\phi^{(Xf)
}_{SL}|\lsim{\cal O}(\tan^2\theta_C)$$ for the phase difference $
\phi^{(Xf)}_{SL}$ in expression (29). The largest value of $\phi^{
(Xf)}_{SL}$, admissible by this estimate, would be achieved at the
relative phase of $a^{(X)}_{DK}$ and $a^{(X)}_{D\overline K}$ equal
to $\pm\pi/2$. Considerations based on final state interactions
tend to change this phase and reduce $\phi^{(Xf)}_{SL}$ even
stronger.

For the quantity $\alpha_D$, appearing in single-time distributions
together with $\phi^{(Xf)}_{SL}$ (see expression (29)), the Minimal
Standard Model gives very small expected values, typically $< 10^{
-3}$. However, various hypotheses on New Physics may lead to larger
values, up to $|\tan\alpha_D|=|x_D|\sim0.1$. The present
experimental data~\cite{pdt} still give a rather weak limitation
$$|\tan\alpha_D|<0.09\,,$$ and cannot exclude such New Physics. So
$\alpha_D$ could be of the same order as $\phi^{(Xf)}_{SL}$ or even
higher, and the problem of their separation looks serious.

Distributions on the secondary-decay time $t_K$ for decays (25)
cannot, by themselves, separate $\phi^{(Xf)}_{SL}$ and $\alpha_D$.
This means that they cannot determine amplitudes $a^{(X)}_{DK}$
and $a^{(X)}_{D\overline K}$ and/or mass difference $\Delta m_D$
(the same is true for using secondary decays of only $K_S$ or $K_L$,
or for total yields of decay products integrated over $t_K$). One
could, however, try to interpret experimental results by applying
additional hypotheses which should be checked. If, for instance,
$$|\alpha_D|\gg|\phi^{(Xf)}_{SL}|\,,$$ then the constant
phaseshifts for oscillating terms in all decays (25) should be the
same; $|\tan\alpha_D|$ should coincide with $|x_D|$ measured in
semileptonic decays of neutral $D$-mesons (note that semileptonic
decays are insensitive to the sign of $x_D$). If $$|\alpha_D|\ll
|\phi^{(Xf)}_{SL}|\,,$$ then useful would be comparison with
similar phase shifts in the charged $D$-meson decays (30), having
no mixing (these decays are of independent interest as well). Also
useful could be studies of decays of neutral or charged $D$-mesons
to charged kaons. They measure absolute values of amplitudes
isotopically related to amplitudes of transitions (2), (3) for
decays (25).

More accurate separation between different interference effects,
mixing and/or suppressed vs. favored transitions, can be achieved
only by invoking information on double-time distributions. Of
course, detailed studies of double-time oscillations require very
high experimental statistics. One can imagine, however, that they
would not be necessary. For example, comparison of
$t_K$-distributions in various $t_D$-regions (say, $t_D\lsim\tau_D$
and $t_D\gsim\tau_D$) could be sufficient at relatively moderate
statistics. Treatment of the corresponding results could be
simplified by taking into account the smallness of $\Delta m_D$
and $\Delta\Gamma_D$. To achieve more definite judgment on
experimental availability of such studies one needs various Monte
Carlo simulations. In any case, necessary $CP$-parities of heavier
and lighter $D$-eigenstates (if they are not mode-dependent) could
be measured in special experiments, and used afterwards in all
other studies (just as done for kaons).

In summary, we have shown that the phenomenon of coherent
double-flavor oscillations (CDFO) in cascade decays of heavy
neutral flavored mesons into intermediate neutral kaons is very
useful to study heavy mesons and their decays.  The phenomenon
reveals itself mainly in double-time decay distributions (over
the primary and secondary decay times). It gives, first of all,
possibility to determine $CP$-parities (exact or approximate) of
the heavy meson eigenstates, suggests new approaches to
investigation of $CP$-violation and (especially for $D$-mesons)
of suppressed flavor-transition amplitudes. It could also check
consistency of various assumptions on the mesons.

On the other hand, CDFO appears to be inevitable to solve some
problems unambiguously and in a model-independent way. They are,
in particular, such important problems as the unambiguous
measurement of $CP$-violating parameters and/or relation of
the meson eigenwidths and eigenmasses. Another problem, specific
for $D$-mesons, is study of doubly Cabibbo-suppressed  transitions
which are coherent with Cabibbo-favored transitions in decays
(of both neutral and charged $D$-mesons) to final states with
neutral kaons. Such studies are very interesting by itself and
may give evidence for New Physics, independent of (and additional
to) $CP$-violation studies. Our main point is that extraction of
both suppressed amplitudes and $CP$-violating parameters for
neutral $D$-mesons appears impossible without investigation of
CDFO. To separate effects of $D$-meson mixing and interference of
suppressed vs. favored amplitudes, such investigations for
$D$-mesons, in contrast to $B$-mesons, require to know double-time
decay distributions (i.e,. over both primary and secondary decay
times). For charged $D$-mesons one should also measure the
secondary decay time distributions to achieve unambiguous
extraction of suppressed amplitude in decays with neutral kaon
production.

\section*{Acknowledgements}

The author is grateful to I.Dunietz and R.M.Ryndin for helpful
and stimulating discussions. The work was supported in part by
the RFBR grant 96-15-96764.

\end{document}